\documentclass[]{aa}
\usepackage{psfig}

\begin{document}

\title{Photometric observations of 9 Near-Earth Objects\thanks{Based
on observations taken at the German-Spanish
Astronomical Centre, Calar Alto, operated by the Max-Planck-Institute for
Astronomy, Heidelberg, jointly with the Spanish National Commission
for Astronomy}}

\author{Gy. Szab\'o\inst{1,4} \and B. Cs\'ak\inst{2,4} \and
K. S\'arneczky\inst{3,4} \and L.L. Kiss\inst{1,4}}

\institute{
Department of Experimental Physics \& Astronomical Observatory,
University of Szeged,
H-6720 Szeged, D\'om t\'er 9., Hungary \and
Department of Optics \& Quantum Electronics \& Astronomical
Observatory, University of Szeged,
POB 406, H-6701 Szeged, Hungary \and
Department of Physical Geography, ELTE University, H-1088 Budapest,
Ludovika t\'er 2., Hungary \and
Guest Observer at Konkoly Observatory}

\titlerunning{Photometric observations of 9 NEOs}
\authorrunning{Szab\'o et al.}
\offprints{l.kiss@physx.u-szeged.hu}
\date{}

\abstract{
We present new CCD observations of nine Near-Earth
Asteroids carried out between February, 1999 and July, 2000.
The bulk of the data was acquired through an R$_{\rm C}$ filter, while
the minor planet 11405 was observed without filter.
We could determine synodic periods and amplitudes for 5
asteroids, 
699: $3\fh3$, $0\fm18$;
1866: $2\fh7$, $0\fm12$;
1999 JD6: $7\fh68$, $1\fm2$;
2000 GK137: $4\fh84$, $0\fm27$;
2000 NM: $9\fh24$, $0\fm30$. 
Based on observations taken at different phases, we could
infer a phase parameter m of 0.018$\pm$0.005 for 1865 Cerberus.
An epoch-method yielded a sidereal period 
of $0\fd27024003(5)$ for this object with retrograde rotation.
The remaining 3 objects have only partial coverage, thus
no firm conclusion on their synodic period is possible.
}

\maketitle

\section{Introduction}

The new observing strategies using fully automatic telescopes
(e.g. project NEAT, Pravdo et al. 1999, LINEAR, 
Stokes et al. 2000) and dedicated large instruments
led to regular discoveries of a wealth of relatively bright asteroids,
including a considerable number of Near-Earth Objects (NEOs).
The quickly responding networks of amateur and
professional observatories allow an accurate determination of
the orbit even for NEOs with high daily motions, avoiding their
loss due to the poor astrometry that was typical a few years ago
(Steel \& Marsden 1996).
Therefore, in solar system studies this is one of the
research fields in which most progress is being made.
During the recent years
there have been intensive studies of these objects owing to
the recognition of their possible impact hazards (Binzel 2000).

The well-covered NEOs can be
studied with photometric methods with great accuracy.
Ground-based shape and rotation models require precise
time-resolved photometry through several oppositions and
this is the main reason why relatively few models in the literature
do exist. Possible binarity or precession similarly affect the
lightcurves, therefore, to choose between the possibilities,
further high-precision lightcurves are needed (see, e.g.,
Binzel 1985, Kiss et al. 1999, Pravec et al. 2000).

We started a long-term observational project with the main
goal of obtaining
CCD photometric observations of selected minor planets
(S\'arneczky et al. 1999, Kiss et al. 1999). This paper
reports new observations of NEOs carried out at three
observatories, between February, 1999 and July, 2000.
The data acquisition, main geometric parameters and
applied methods are described in Sect.\ 2, while
Sect.\ 3 deals with the detailed observational results.

\section{Observations and modeling methods}

\subsection{Data acquisition}

Observations have been made at three different observing sites.
Johnson V and R$_{\rm C}$ filtered CCD observations were carried
out at Calar Alto Observatory (Spain) on 10 nights in June--July, 2000
(we also observed distant active comets during this observing
run -- see Szab\'o et al., 2001).
The instrument used was the 123/981 cm Cassegrain telescope (hereafter
designed by [1]) equipped with the SITe\#2b CCD
camera (2048$\times$2048 pixels giving an angular
unbinned resolution of 0\farcs49/pixel). The projected
sky area is 16\farcm0$\times$16\farcm0, 10\farcm0$\times$10\farcm0
unvignetted. 

We carried out $R_{\rm C}$ filtered CCD observations at Piszk\'estet\H o
Station of Konkoly Observatory on two nights in September, 1999
and on one night in January, 2000.
The data were obtained using the 60/90/180~cm
Schmidt-telescope [2] equipped with a Photometrics AT200 CCD
camera (1536$\times$1024 KAF 1600 MCII coated CCD chip). The projected
sky area is 29$^\prime$$\times$18$^\prime$ which corresponds
to an angular resolution
of 1\farcs1/pixel. The photometric effect of neglecting standard
transformation
was discussed in Kiss et al. (1999), where we illustrated the usability
of single-filtered data.

R-filtered and unfiltered CCD photometry was carried out on two
nights at the University of Szeged using a
0.28-m Schmidt-Cassegrain telescope [3] located in
the very center of the city of Szeged. Two detectors were used.
The SBIG ST-6 CCD camera (375$\times$242 pixels) gives an angular
resolution of about 2 $^{\prime\prime}$/pixel (the pixels are rectangular).
The second detector was
an SBIG ST-7 CCD camera (765$\times$510 pixels) giving an angular
resolution of about 1\farcs0/pixel. The pixels are also rectangular.

The instruments used together with the times of exposure are summarized
in Table 1. 
The last column gives the comparison stars.
In the case of relative photometry the local comparison stars in the
fields are identified by their GSC 1.2 (Guide CD-ROM Star Chart, 1997) 
or USNO A2.0 (Monet et al. 1998) numbers and magnitudes. When
absolute photometry was done (at Calar Alto Observatory), 
the standard fields of Landolt (1992)
are included.
Because of the fast motion of 1999 CV3, the apparent path of the
asteroid could be covered only by three overlapping frames, therefore,
three different comparison stars were used on the same night.
The aspect data of the observations are presented in Table 2.

\begin{table*}
\caption{Data of the used instruments and comparison stars. 
Subscripts 'i' and 's' denote whether the data are in
instrumental or standard photometric system.
See text for telescope codes.}
\begin{center}
\begin{tabular} {llllll}
\hline
Date & Telescope
& Detector & Filter & Exp (s) & Comp. \\
\hline
{\bf 699 Hela} & & & &&\\
1999 09 15/16 & [3] & ST7 & R$_i$ & 50 & GSC 2734 1730 \\
{\bf 1865 Cerberus} & & & & &\\
1999 09 24/25 & [2]& AT200& R$_i$ & 60 & USNO 012243+254115\\
2000 07 08/09 & [1] & SITe\#2b & R$_s$ &240& PG 1633+099\\
2000 07 09/10 & [1] & SITe\#2b & R$_s$ &240& PG 1633+099\\
2000 07 10/11 & [1] & SITe\#2b & R$_s$ &240& PG 1633+099\\
{\bf 1866 Sisyphus} & & & & &\\
2000 06 30/07 01 & [1] & SITe\#2b & R$_s$ &240& PG 1525-071\\
{\bf (11405) 1999 CV3} & & & & &\\
1999 02 27.8 & [3] & ST6 & unf & 30 &GSC 3434 172\\
1999 02 27.9 & [3] & ST6 & unf & 30 &GSC 3434 144\\
1999 02 28.0 & [3] & ST6 & unf & 30 &GSC 3434 96\\
{\bf (16064) 1999 RH27} & & & & &\\
2000 01 01/02 & [2] & AT200 & R$_i$ & 60 & USNO 043039+395018 \\
{\bf 1999 JD6} & & & & &\\
2000 07 02/03 & [1] & SITe\#2b & R$_s$ &60& PG 1633+099\\
2000 07 05/06 & [1] & SITe\#2b & R$_s$ &60& PG 1633+099\\
{\bf 1999 ND43} & & & &&\\
1999 09 23/24 & [2] & AT200 & R$_i$ & 60& GSC 3302 711\\
{\bf 2000 GK137} & & & & &\\
2000 06 29/30 & [1] & SITe\#2b & R$_s$ &60& PG 1657+078\\
2000 07 01/02 & [1] & SITe\#2b & R$_s$ &60& PG 1657+078\\
{\bf 2000 NM} & & & & &\\
2000 07 03/04 & [1] & SITe\#2b & R$_s$ &20& S.Area 110\\
2000 07 04/05 & [1] & SITe\#2b & R$_s$ &20& S.Area 110\\
2000 07 05/06 & [1] & SITe\#2b & R$_s$ &20& S.Area 110\\
\hline
\end{tabular}
\end{center}
\end{table*}

\begin{table*}
\caption{The journal of observations. ($r$ -- heliocentric distance;
$\Delta$ -- geocentric distance; $\lambda$ -- ecliptic longitude;
$\beta$ -- ecliptic latitude; $\alpha$ -- solar phase angle;
aspect data are referred to 2000.0)}
\begin{center}
\begin{tabular} {llrrlrrr}
\hline
Date & RA & Decl. & $r$(AU) & $\Delta$(AU) & $\lambda (^\circ)$ & $\beta 
(^\circ)$ & $\alpha (^\circ)$\\
\hline
{\bf 699 Hela} & & & & & & & \\
1999 09 15/16 & 22 28 & +30 56 & 1.57 & 0.65 & 327 & 20 & 23\\
{\bf 1865 Cerberus} & & & & & & &\\
1999 09 24/25 & 01 22 & +25 38 & 1.55 & 0.61  & 28 & 16 & 20 \\
2000 07 08/09 & 00 11 & +27 41 & 1.37 & 0.87 & 9 & 28 & 48\\
2000 07 09/10 & 00 11 & +27 59 & 1.37 & 0.87 & 9 & 28 & 48\\
2000 07 10/11 & 00 12 & +28 05 & 1.38 & 0.86 & 10 & 28 & 47\\
{\bf 1866 Sisyphus} & & & & \\
2000 06 30/07 01 & 12 49 & 26 57 & 2.73 & 2.69 & 195 & 20 & 22\\
{\bf (11405) 1999 CV3} & & & & \\
1999 02 27.8 & 10 00 & +46 48 & 1.16 & 0.22 & 168 & 53 & 32.7\\
1999 02 27.9 & 10 00 & +47 02 & 1.16 & 0.22 & 168 & 53 & 33.0\\
1999 02 28.0 & 10 00 & +47 15 & 1.16 & 0.22 & 168 & 53 & 33.2\\
{\bf (16064) 1999 RH27} & & & & \\
2000 01 01/02 & 04 31 & +39 48 & 1.24 & 0.29 & 54 & 61 & 26\\
{\bf 1999 JD6} & & & & \\
2000 07 02/03 & 17 48 & +10 30 & 1.34 & 0.38 & 263 & $-$13 & 28\\
2000 07 05/06 & 17 33 & +10 45 & 1.32 & 0.37 & 261 & $-$13 & 30\\
{\bf 1999 ND43} & & & &\\
1999 09 23/24 & 02 17& +49 21 & 1.06 &0.10 & 50 & 33 &53\\
{\bf 2000 GK137} & & & & \\
2000 06 29/30 & 19 55 & +58 33 & 1.05 & 0.17 & 287 & 37 & 74\\
2000 07 01/02 & 20 19 & +60 34 & 1.04 & 0.17 & 294 & 39 & 77\\
{\bf 2000 NM} & & & & \\
2000 07 03/04 & 18 09 & -05 06 & 1.17 & 0.16 & 272 & $-$5 & 18\\
2000 07 04/05 & 18 07 & -01 59 & 1.16 & 0.16 & 272 & $-$2 & 21\\
2000 07 05/06 & 18 04 & +01 21 & 1.15 & 0.15 & 272 & 0 & 24\\
\hline
\end{tabular}
\end{center}
\end{table*}

The exposure times were limited by two factors: firstly, the
asteroids were not allowed to move more than the
FWHM of the stellar profiles (varying from night to night) and
secondly, the signal-to-noise (SN) ratio had to be at least 10.
This latter parameter was estimated by comparing the peak pixel values
with the sky background during the observations.

\begin{figure}
\begin{center}
\leavevmode
\psfig{figure=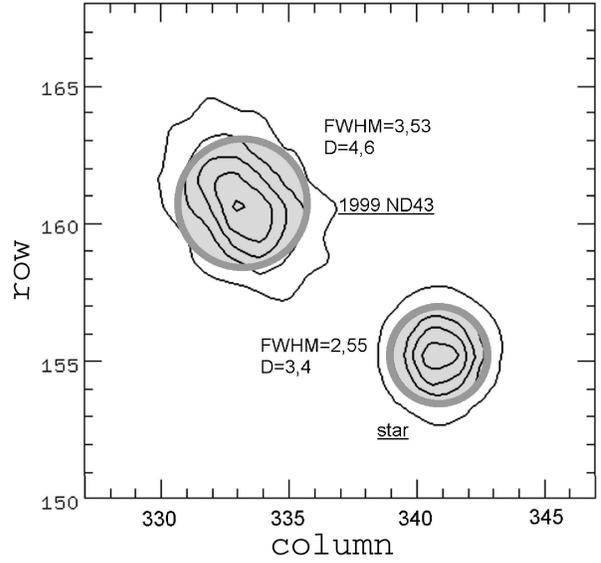,width=8cm}
\caption{Profiles of a star and an asteroid fitted with the optimal
aperture of the reduction (see text for details).}
\end{center}
\label{f1}
\end{figure}

Selecting the appropriate size of 
the photometric aperture is of considerable importance
when doing aperture-photometry of faint objects with remarkable
fast motion. Selecting too small apertures causes loss of light,
while selecting too large radii increases the noise in the
aperture. Usually, observers select a series of apertures during
the reduction, and the ``best'' lightcurve is accepted. 
In our analysis a different approach has been used. 
As the elongated profiles of the
moving asteroids can be hardly treated analytically,
we have performed a semi-experimental optimalization
of the signal-to-noise ratio based on the
photometric behaviour of the profiles.

Assuming a homogeneous noise model with the background scatter density value
(scatter/picture element) {\it n}, the square of the total noise
{\it N} present in the aperture with a radius of {\it R} can be 
taken as the sum of the noise in every picture element:

\begin{equation}
$$N^2 = n^2 ({R^2 \pi}) = (1.772R\cdot n)^2$$.
\end{equation}

Let $S$ be the measured signal. The aperture radius 
with the highest S/N ratio is established by taking its differential

\begin{eqnarray}
\lefteqn{d{S \over N} = {N \cdot dS - S \cdot dN \over N^2} =
1.772n {R \cdot dS - S \cdot dR \over (1.772 \cdot R) ^2}=}\nonumber \\
\lefteqn{= {n \over 1.772R^2}(R \cdot dS-S \cdot dR)= 0}
\end{eqnarray}

As the whole expression equals 0 for the optimal $S/N$,
one can divide the
equation by ${n \over 1.772R^2}$. 
Elimination of $n$ yields to the fundamental conclusion
that the diameter of the optimal aperture does not depend on the
scatter of the background. That is true for all circular and
elongated profiles on the assumption of homogeneous noise.

After rearranging Eq. (2) one gets $RdS=SdR$, the solution of which is:

\begin{equation}
$${\rm ln}~R = {\rm ln}~S = {\rm ln}~ 10^{-0.4*(m-m_0)} = -0.92(m-m_0)$$,
\end{equation}

where $m$ is the measured brightness in the aperture, expressed
in magnitudes, while $m_0$ is the integration constant.
In an aperture of radius $R+\Delta R$ the measured magnitude
is $m-\Delta m$. If $\Delta m$ is small, the following expression
will be also true for the increased aperture:

\begin{equation}
$${\rm ln}~(R+\Delta R) = -0.92(m-m_0-\Delta m)$$.
\end{equation}

Taking the difference (4)--(3) and using 
${\rm ln}~(R+\Delta R)-{\rm ln}~(R) = \Delta R/R$,
we get the final expression valid for the optimal aperture:

\begin{equation}
$${\Delta R \over R} = 0.92 \Delta m$$.
\end{equation}

The left-hand side expresses the relative increment of the radius, while
the right-hand side is the corresponding magnitude change.
In practice, the radius of the aperture is altered and the alteration
of the measured magnitude is examined. When their values satisfy
Eq. (5), the aperture is accepted to be the optimal one.

The projection of the calculated aperture onto the stellar and
asteroidal profiles is presented in Fig. 1. We note
that the fit of the circular aperture with the elongated profile
is not very good, which
decreases the attainable S/N ratio.
The optimal aperture radius depends slightly on the
apparent motion, its average size is about 130 percent of the FWHM.

The image reduction was done with standard IRAF routines.
Aperture photometry 
was performed with the IRAF task {\it noao.digiphot.apphot.qphot}.
Two nearby comparison stars were selected and their
differential light curves were examined to estimate the photometric 
accuracy. The scatter
arising from the comparison stars was estimated to be $\pm0\fm01$--$0\fm03$,
depending on the weather conditions. (We note, that our time series 
observations were also analysed by Cs\'ak et al. 2000 resulting in 
the discovery of 13 new short period variable stars. Therefore,
we could solidly exclude variables from the chosen comparison stars.)
The photometric stability 
indicated by the relatively low scatter of 
the brightness differences of the comparison stars
suggests that the motion and
the faintness of the asteroid is the dominating factor for
the higher scatter of the lightcurves. Thus, 
the above described optimalization procedure was 
applied for the detected asteroid profiles resulting in 
a slightly larger aperture radius than that for the 
stellar profiles, namely $1.3\times$FWHM. The comparison stars were, 
of course, measured with the same aperture.

In the case of the instrumental lightcurves,
the presented magnitudes throughout the paper are based on magnitudes
of the comparison stars taken from the Guide Star Catalogue (GSC) (Table 1).
Therefore, their absolute values are fairly uncertain
(at a level of $\pm$0\fm2--0\fm3). Fortunately, the shape of the
lightcurve is not affected by neglecting the standard transformations.

The sky conditions at Calar Alto Observatory allowed doing
standard all-sky photometry. The standard sequences were selected 
from the list of Landolt (1992), containing 4-5 standard stars covering 
V magnitudes between
13--15 magnitudes and V$-$R indices between $-$0\fm1 and 0\fm6.
In order to monitor the extinction correction, the fields were 
observed in every hour.

The presented data were corrected for light-travel time
and are available electronically at CDS-Strasbourg.

\subsection{Methods of analysis}

The main aim of our observations was to get information on the
rotational states of NEOs. The period of rotation can be inferred
when more than one rotational phase is covered. Determination
the sense of rotation requires a more complex analysis of lightcurves
taken at different ecliptic positions. 
As there have been sufficient data series 
in the literature for 1865 Cerberus, a rotation model
could be also calculated. 

Two main types of method are used widely. The amplitude-method
was described, e.g., by Magnusson (1989). For this,
the amplitude information is used to determine the spin vector and the
shape. An important point is that the observed $A(\alpha)$
amplitudes at solar
phase $\alpha$ should be reduced to zero phase ($A(0^\circ)$), if possible,
by a simple linear transformation
in form of $A(\alpha)=A(0^\circ)(1+m \alpha)$. 

The other possibility is to examine the times of light extrema
(``epoch-methods'', ``E-methods'', see, e.g., Magnusson 1989,
Detal et al. 1994).
In this paper a modified version is used
by constructing the O$-$C'--ecliptic longitude diagrams as a 
function of revolution. Examining the virtual phase shifts 
(``observed minus calculated'') occurring
during a revolution helps to establish the sidereal period of rotation.
The trend of the phase shifts gives the sense of rotation.
Increasing O$-$C' (i.e. phase shift) refers to prograde rotation,
while retrograde rotators have decreasing O$-$C'.
The method is briefly outlined in Kiss et al. (1999) and in Szab\'o et al.
(1999), and will be thoroughly presented and discussed in 
a forthcoming paper (Szab\'o et al., in prep.).

\section{Discussion}

\noindent
In the following, we describe the observational details
and results.

\bigskip
\noindent {\it 699 Hela}

\noindent This asteroid was discovered by J. Helffric in
Heidelberg, in 1910. Although this unusual object does not belong to
the real Earth-grazing asteroids, it can pass quite close to 
the Earth in perihelion
(q=1.55 AU).
The first photometry was done by Binzel (1987), who presented
nine data points. He obtained a period of about 3 hours, while
the measured amplitude was over 0\fm53 at the ecliptic position
$\lambda=300^\circ$ and 
$\beta=18\fdg8$ (the solar phase angle was $\alpha=23^\circ$).

Our observations carried out with instrument [3] on September 15, 1999
were done very close to the perihelion passage. The amplitude
of 0\fm18 turned out to be smaller than the one previously detected.
As the solar phase angle is similar for the two sets of observations
(ours and that of Binzel 1987),
the significant difference of the amplitudes cannot be the effect
of the varying aspect angle. 
The most likely value for the synodic period is 
3\fh3$\pm$0\fh2.
The lightcurve is presented in Fig. 2.

\begin{figure}
\begin{center}
\leavevmode
\psfig{figure=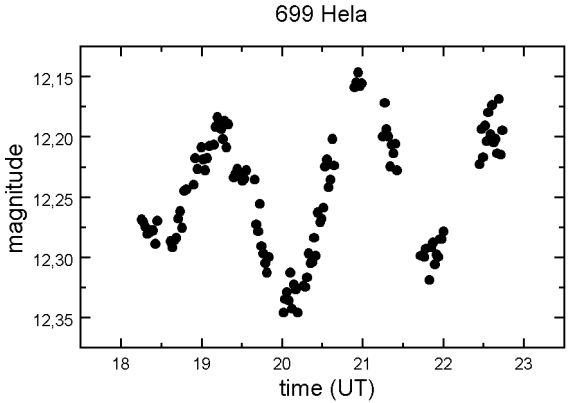,width=8cm}
\caption{The $R_{\rm C}$ lightcurve of 699 on September 15th, 1999.}
\end{center}
\label{f2}
\end{figure}

\bigskip
\noindent {\it 1865 Cerberus}

\noindent This asteroid has one of the largest
observed amplitudes of all minor planets. The first photometry including
41 data points
was discussed in Harris \& Young (1989); later, Wisniewski et al. (1997)
presented the second lightcurve of this asteroid.
We presented the lightcurve in the 1998 opposition in S\'arneczky et al.
(1999).
In this paper both the lightcurves of the opposition in 
1999 (Fig.\ 3) and the composite lightcurve
observed between 08/09--10/11 July (Fig 4.), 2000 are presented.
Measured amplitudes (A$_{\rm obs}$) and times of extrema (t$_{min}$)
at different aspects are summarized in Table 3.

\begin{table*}
\begin{center}
\caption{Published photometries of 1865 Cerberus}
\begin{tabular} {lrrrlll}
\hline
Date & $\lambda (^\circ)$ & $\beta (^\circ)$ & $\alpha (^\circ)$ & A$_{\rm 
obs}$ & t$_{min}$ & ref. \\
\hline
1980 11 04-06    &  40   &   $-$12   &  10  &  1\fm45  & 2444548.979 & (1) \\
1989 11 03-04  &  63   &   $-$17   &  22  &   1.8    & 2447835.76245 & (2)\\
1998 10 23,26  &  42   &   $-$2.8  & 38.7 &   2.3    & 2451113.515 & (3)\\
1999 09 24     &  28    &   16      & 20   &   1.7    & 2451446.455 & (4)\\
2000 07 08-10  &   9   &   28      & 48   &   2.2    & 2451734.536 & (4)\\
\hline
\end{tabular}
\end{center}
References: (1) -- Harris \& Young 1989; (2) -- Wisniewski et al.
1997; (3) -- S\'arneczky et al. 1999;
(4) -- present paper
\end{table*}

Based on earlier data (see Table 3), we tried to determine a new
model by the amplitude method.
The $m$ parameter was estimated by minimizing the
scatter of the longitude--amplitude diagram,
resulting $m=0.018\pm 0.005$.

The O$-$C' model has been determined (Fig.\ 5). Because of
the distribution of data, the spin axis cannot be
estimated based on the O$-$C' points.
The ecliptic position was almost the same at the time of
the observations in 1980 and 1988,
though the observations were made many years apart.
This has quite a favourable influence on the calculations
of the sidereal period.
A value of $P_{\rm sid}$=0\fd27024003$\pm$0\fd00000005 is resulted. 
The O$-$C' graph decreases which indicates retrograde rotation.

\begin{figure}
\begin{center}
\leavevmode
\psfig{figure=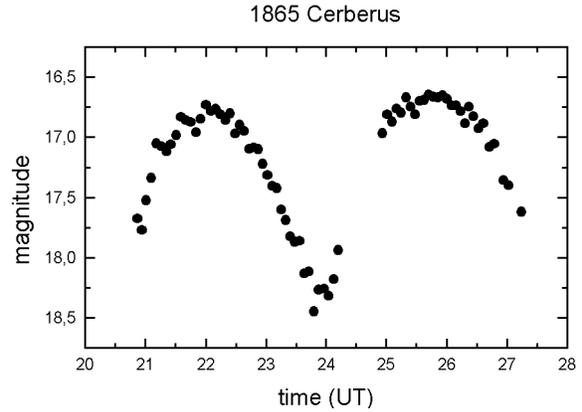,width=8cm}
\caption{The instrumental $R_{\rm C}$ lightcurve of 1865 on September
24th, 1999.}
\end{center}
\label{f3}
\end{figure}

\begin{figure}
\begin{center}
\leavevmode
\psfig{figure=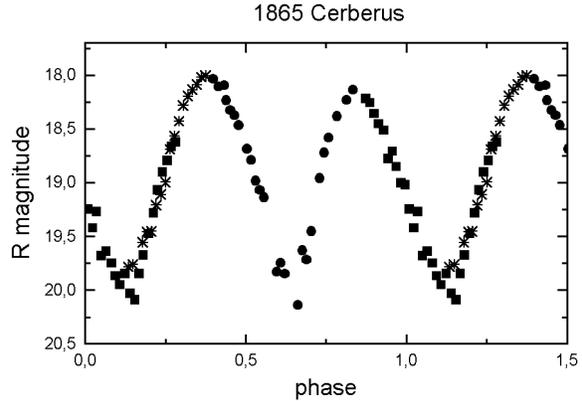,width=8cm}
\caption{The composite $Rc$ lightcurve of 1865 in July, 2000. 
Data on the individual nights are denoted by
solid squares (07.08), solid circles (07.09) and asterisks (07.10). 
Zero phase is at JD 2451734.490.}
\end{center}
\label{f4}
\end{figure}

\begin{figure}
\begin{center}
\leavevmode
\psfig{figure=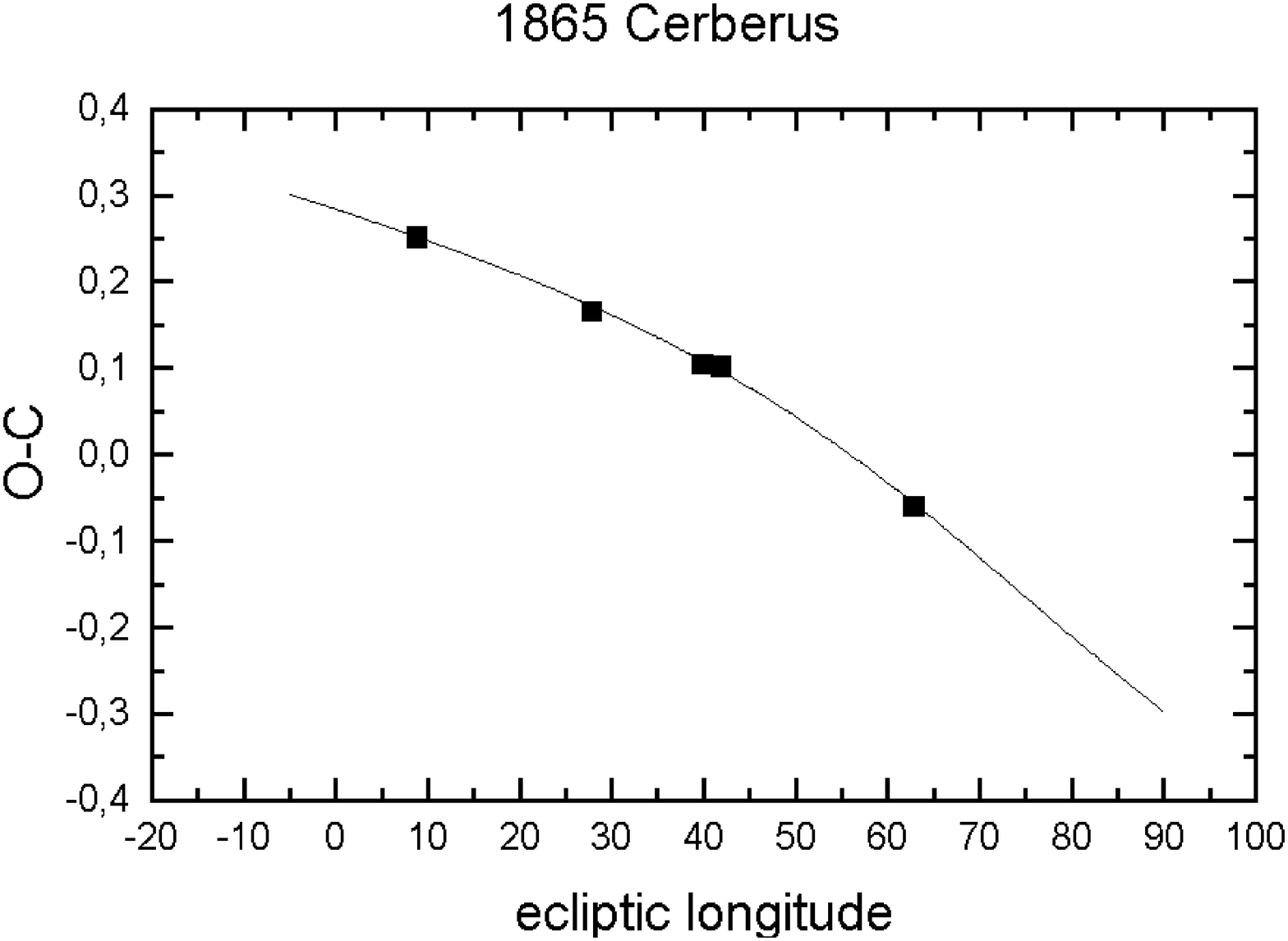,width=8cm}
\caption{The observed O$-$C' values fitted with the model for 1865 Cerberus.}
\end{center}
\label{f5}
\end{figure}

\bigskip
\noindent {\it 1866 Sisyphus}

\noindent
This asteroid was discovered by P. Wild using a 40 cm Schmidt-telescope
in 1972. The first photometry was presented by Schober et al. (1993).

\begin{figure}
\begin{center}
\leavevmode
\psfig{figure=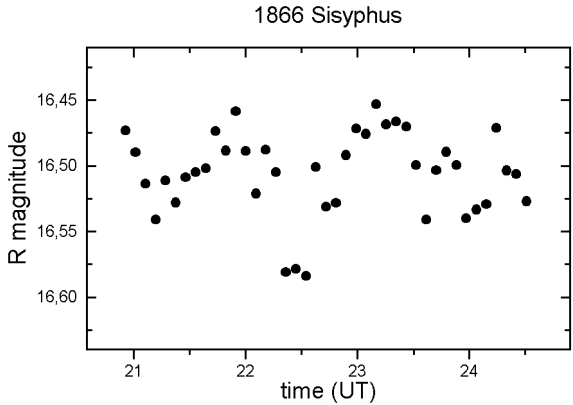,width=8cm}
\caption{The $R_{\rm C}$ lightcurve of 1866 on June 30th, 2000.}
\end{center}
\label{f6}
\end{figure}

During our observations the previously measured period of 2.7 hours
was confirmed. The amplitude was 0\fm12 close to
the previously determined 0\fm13 (Schober et al., 1993).
During the observations the seeing was about 0\farcs7 at the
altitude of the asteroid. This circumstance 
had a fortunate influence on the
data reduction, although the apparent motion was about 
1\farcs6 during the 4 minutes of integration. 
The lightcurve is presented in Fig.\ 6.

\bigskip
\noindent {\it (11405) 1999 CV3}

\noindent The asteroid was discovered by the LINEAR project
in Socorro. Thanks to its close path to the Earth
its maximum brightness was about 13 mag on the night of the observation.
Its proper motion was quite fast with a value of 360$^{\prime\prime}$/hour,
which is why we had to use 3 different comparison
stars located on 3 overlapping star fields along the path during the 5 hours
of observation. The fast motion did not permit exposure times of 
over 20 seconds.

Its period is longer than the observing interval but assuming a simple
``double'' lightcurve it may be about 7 hours. The measured amplitude is
0\fm43. The lightcurve is presented in Fig. 7.

\begin{figure}
\begin{center}
\leavevmode
\psfig{figure=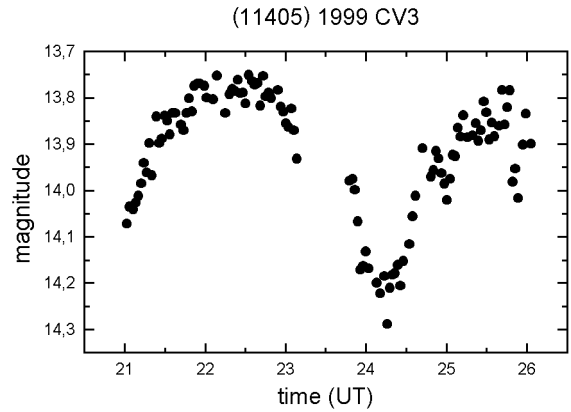,width=8cm}
\caption{The unfiltered lightcurve of 1999 CV3 on February 27th, 1999.}
\end{center}
\label{f7}
\end{figure}

\bigskip
\noindent {\it (16064) 1999 RH27}

This Earth-grazing asteroid was discovered by the Catalina
Sky Survey on 5th September, 1999. During our observations
at Konkoly Observatory, a quite
ambiguous light variation was detected. During the 3 hours
of observation the lightcurve showed a decrease of 0\fm1.
The amplitude is higher than this value while the period seems to be
much longer than the observing run. The lightcurve is
presented in Fig. 8.

\begin{figure}
\begin{center}
\leavevmode
\psfig{figure=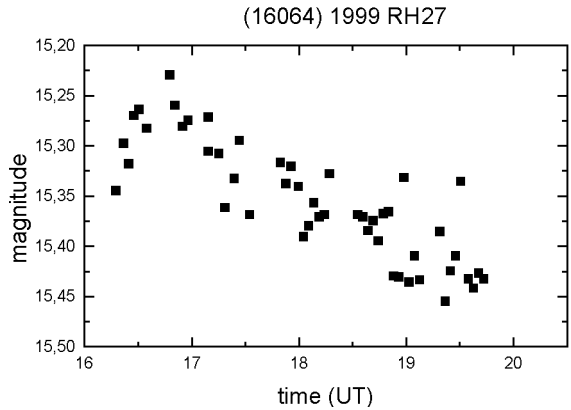,width=8cm}
\caption{The $R_{\rm C}$ lightcurve of 1999 RH27 on January 1st, 2000.}
\end{center}
\label{f8}
\end{figure}

\bigskip
\noindent{\it 1999 JD6}

\noindent This member of the {\it Aten} family was discovered
by the LONEOS project in Flagstaff. The equipment used was a
59 cm Schmidt-telescope. It turned out to be another high-amplitude
NEO with a value of 1\fm2$\pm$0\fm1 magnitude. The period of the
light variation is 7\fh68$\pm$0\fh03 hours.
The lightcurve is quite similar to the other high-amplitude
asteroids, e.g. 1865 Cerberus as the maxima take much more time
than the sharp minima. That can be explained by the rotation of
a splinter-like body. The composite lightcurve is presented in
Fig.\ 9.

\begin{figure}
\begin{center}
\leavevmode
\psfig{figure=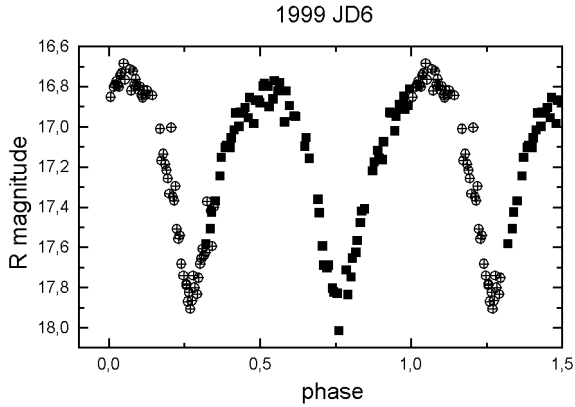,width=8cm}
\caption{Composite lightcurve of 1999 JD6.
Data series of the individual nights are denoted by
crossed circles (07.02) and solid squares (07.05).
Zero phase is at 2451728.38408.}
\end{center}
\label{f9}
\end{figure}

\bigskip
\noindent{\it 1999 ND43}

\noindent This asteroid was discovered also by LINEAR in Socorro,
on July 14th, 1999. The instrument used was the 1 meter reflector.

During the observations the asteroid was in a distance of
0.10 AU. Its apparent motion was almost 3$^{\prime\prime}$/min, 
but because of
its faintness we had to increase the exposure time up to 1 minute.
The highly elongated profile was presented in Fig. 1. in Sect. 2.
The 5 hours-long lightcurve does not show periodic light variation,
only an obvious faintening of about 0\fm5 is visible, as 
presented in Fig. 10.

\begin{figure}
\begin{center}
\leavevmode
\psfig{figure=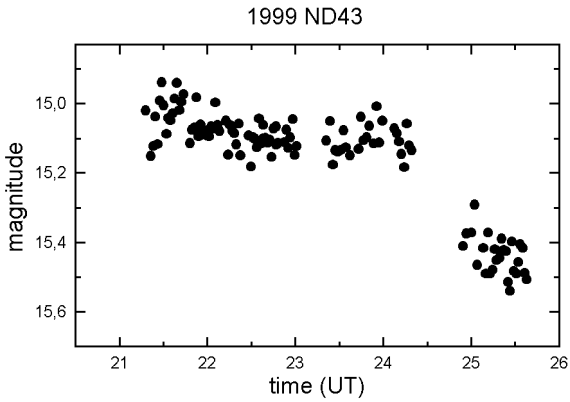,width=8cm}
\caption{The $R_{\rm C}$ lightcurve of 1999 ND43 on September 23th, 1999.}
\end{center}
\label{f10}
\end{figure}

\bigskip
\noindent{\it 2000 GK137}

This NEO was discovered by the LINEAR project in Socorro. The object was
included in our target list following the suggestions of P. Pravec
(personal communication).
In the present paper, data obtained on two nights are presented.
The composite lightcurve was calculated using a period of
4\fh84$\pm$0\fh02. 
The shape of the lightcurve is quite asymmetric, the brighter
maximum is followed by a nearly constant segment. Considering the
high solar phase during the observing run ($\alpha\approx75^{\circ}$),
the lightcurve is likely to simply reflect local surface irregularities.
The total amplitude of the lightcurve is 0\fm27.

\begin{figure}
\begin{center}
\leavevmode
\psfig{figure=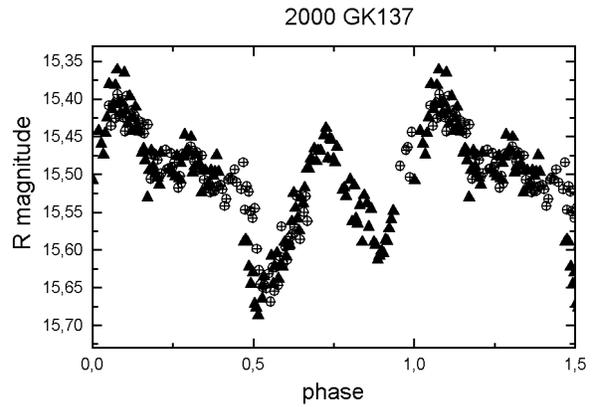,width=8cm}
\caption{Composite lightcurve of 2000 GK137.
Data series of the individual nights are denoted by
crossed circles (06.29) and up triangles (07.01).
Zero phase is at 2451727.37817.}
\end{center}
\label{f11}
\end{figure}

\bigskip
\noindent{\it 2000 NM}

\noindent This asteroid was discovered by 
amateur astronomer L.L. Amburgey July 2.17 UT, 2000. The
instrument used was a 21 cm telescope and the asteroid was found
on CCD images at a brightness of 13 magnitudes. 39 hours after
its discovery we observed the light variation.
Because of the fast motion the exposure time had to be decreased
to 20 seconds which was also permitted by the high brightness
of the asteroid.
Based on the observations obtained on three nights the period 
has been
determined to be 9\fh24 $\pm$ 0\fh02, while the amplitude
is 0\fm30. The composite diagram is presented in Fig.\ 12.

The lightcurve has three humps similar to 2000 GK137.
The ratio between the shorter and longer hump is 1:2, but
contrary to 2000 GK137, the shorter maximum has the brighter maximal
brightness. The irregularities in the lightcurve are quite surprising
as they cannot be explained by the effect of the high
solar phase (see Table 1 and the discussion of 2000 GK137).

\begin{figure}
\begin{center}
\leavevmode
\psfig{figure=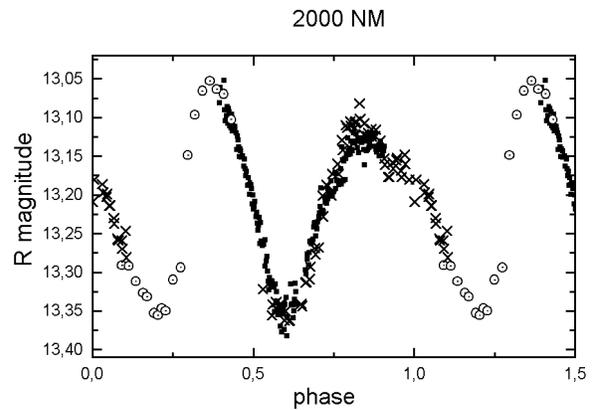,width=8cm}
\caption{The composite lightcurve of 2000 NM.
Data series of the individual nights are denoted by
solid squares (07.03), dotted circles (07.04) and $\times$ (07.05).
Zero phase is at 2451731.50942.}
\end{center}
\label{f12}
\end{figure}

We summarize the resulting synodic periods, amplitudes and models
in Table 4.

\begin{table}
\begin{center}
\caption{The determined periods, amplitudes, spin vectors and shapes.}
\begin{tabular} {llll}
\hline
Asteroid      & P$_{\rm syn} (h)$ & P$_{\rm sid} (d)$ & A$_{\rm obs}$ (mag) \\
\hline
699       & 3\fh3 & & 0\fm18 \\
1865      & &0\fd27024003& \\
1866      & 2\fh7 & & 0\fm12 \\
11405     & $>$5 & & 0.43 \\
16064 & $>$3 & & $>$0.1 \\
1999 JD6 & 7.68 & & 1.2 \\
1999 ND43   & $>$5 & & $>$0.5 \\
2000 GK137 & 4.84 & & 0.27 \\
2000 NM   & 9.24 & & 0.30 \\
\hline
\end{tabular}
\end{center}
\end{table}

\begin{acknowledgements}

This research was supported by the FKFP Grant 0010/2001, 
Pro Renovanda Cultura Hungariae Grants DT 2000. m\'aj./43.,
DT 2000. m\'aj./44., DT 2000. m\'aj./48.
and DT 1999. \'apr./23, OTKA Grant \#T032258, ``Bolyai J\'anos'' Research
Scholarship of LLK from the Hungarian Academy of Sciences
and Szeged Observatory Foundation. The warm hospitality and helpful assistance
of the staff of Calar Alto Observatory and Konkoly Observatory and
their provision of telescope time is gratefully acknowledged.
The NASA ADS Abstract Service was used to access data and references.

\end{acknowledgements}

\end{document}